\documentclass[english,10pt,a4paper]{article}
\usepackage{graphicx,babel,geometry,amssymb,amsmath}
\usepackage[T1]{fontenc}
\usepackage[latin9]{inputenc}

\begin{document}
 
\title{Motion of Bound Domain Walls in a Spin Ladder}
\author{Indrani Bose
        \thanks{indrani@bosemain.boseinst.ac.in} 
        and Amit Kumar Pal}
\date{}

\maketitle
\begin{center}
Department of Physics \linebreak
Bose Institute \linebreak
93/1 A. P. C. Road, Kolkata - 700009 
\end{center}

\medskip{}
\medskip{}

\begin{abstract}
The elementary excitation spectrum of the spin-$\frac{1}{2}$ antiferromagnetic (AFM)
Heisenberg chain is described in terms of a pair of freely propagating spinons. In the 
case of the Ising-like Heisenberg Hamiltonian spinons can be interpreted as domain walls (DWs)
separating degenerate ground states. In dimension $d>1$, the issue of spinons as elementary 
excitations is still unsettled.
In this paper, we study two spin-$\frac{1}{2}$ AFM ladder models
in which the individual chains are described by the Ising-like Heisenberg
Hamiltonian. The rung exchange interactions are assumed to be pure Ising-type in
one case and Ising-like Heisenberg in the other. Using the low-energy
effective Hamiltonian approach in a perturbative formulation, we show
that the spinons are coupled in bound pairs. In the first model, the
bound pairs are delocalized due to a four-spin ring exchange term
in the effective Hamiltonian. The appropriate dynamic structure factor
is calculated and the associated lineshape is found to be almost symmetric
in contrast to the 1d case. In the case of the second model, the bound
pair of spinons lowers its kinetic energy by propagating between chains.
The results obtained are consistent with recent theoretical studies
and experimental observations on ladder-like materials.
\end{abstract}

\medskip{}
\medskip{}

\section{Introduction}

Low-dimensional quantum antiferromagnets exhibit a rich variety of
phenomena indicative of novel ground and excited state properties
\cite{key-1,key-2,key-3}. In one dimension (1d), the ground and low-lying excited
states of the spin chain, in which nearest neighbour spins of magnitude
$\frac{1}{2}$ interact via the antiferromagnetic (AFM) Heisenberg
exchange interaction, can be determined exactly using the Bethe Ansatz
\cite{key-4,key-5}. The ground state has no long range order and the spin-spin
correlations are characterized by a power-law decay. The elementary
excitation is not the conventional spin-1 magnon but a pair of spin-$\frac{1}{2}$
excitations termed spinons. The physical origin of spinons can be
best understood in the Ising limit of the exchange interaction Hamiltonian
given by

\small{
\begin{eqnarray}
H & = & \sum_{i=1}^{N} J_{z}S_{i}^{z}S_{i+1}^{z}
+\sum_{i=1}^{N}\frac{J_{xy}}{2}\left(S_{i}^{+}S_{i+1}^{-}+S_{i}^{-}S_{i+1}^{+}\right)
\end{eqnarray}
}
where $S_{i}^{\pm}$ are the spin raising and lowering operators and
$N$ is the total number of spins. In the Ising limit $(J_{xy}=0)$,
the ground states of the Hamiltonian are the doubly degenerate N\'{e}el
states, one of which is shown in figure 1(a). An excited state is
created by flipping a spin from its ground state arrangement, e.g.,
a down spin is flipped into an up spin in figure 1(b). The flip gives
rise to two domain walls (DWs) consisting of parallel spins and shown
by dotted lines in the figure \cite{key-1,key-6}. The transverse exchange interaction
term in (1) interchanges the spins in an antiparallel spin pair and
has the effect of making the DWs propagate independently (figure 1(c)).
Since the original spin flip carries spin one, each of the DWs or
spinons has spin-$\frac{1}{2}$ associated with it. Spinons are thus
examples of fractional excitations in an interacting spin system.

Spinons can be detected through inelastic neutron scattering in which
neutrons scatter against spins to create spin flips. Due to energy
and momentum conservation, the energy absorption spectrum for spin
flips at different wave vectors can be measured. One observes a peak
at a well-defined energy if the spin flip creates a single particle
excitation. In the case of a pair of spinons, the total energy $\epsilon$
and momentum $k$ of the spin excitation are given by 
$\epsilon(k)=\epsilon_{1}(k_{1})+\epsilon_{2}(k_{2})$
and $k=k_{1}+k_{2}$ where $\epsilon_{i}$, $k_{i}$ $(i=1,2)$ denote
individual spinon energy and momentum \cite{key-1,key-6}. The total momentum
$k$ of the spin flip can be distributed in a continuum of ways among
the spinons giving rise to a continuous absorption spectrum. For a
single particle excitation, the energy versus momentum relation defines
a single branch of excitations whereas for spinons a continuum of
excitations with well-defined lower and upper boundaries is obtained.
The compounds $CsCoCl_{3}$ and $CsCoBr_{3}$ are good examples of
Ising-like Heisenberg antiferromagnets in 1d above the N\'{e}el temperature
and provide evidence of the two-spinon continuum in neutron scattering
experiments \cite{key-6,key-7,key-8}. In the case of the isotropic Heisenberg Hamiltonian
($J_{z}=J_{xy}$ in equation (1)), the spinon spectrum has been clearly
observed in the linear chain compound $KCuF_{3}$ \cite{key-9} though a
physical interpretation of spinons is not as straightforward as in
the Ising-like case.

The existence of spinons, with fractional quantum number spin-$\frac{1}{2}$,
is well established in 1d Heisenberg-type antiferromagnets. In higher
dimensions, the spin-1 magnons are the elementary excitations in magnetically
ordered ground states. There are theoretical suggestions that quantum
antiferromagnets with spin-liquid (no magnetic long range order and
without broken symmetry) ground states may support elementary spinon-like
excitations with fractional quantum numbers \cite{key-10,key-11,key-12}. A well-known
example is that of a resonating-valance-bond (RVB) state, a linear
superposition of VB states, in which the spins are paired in singlet
(VB) configurations. A broken VB gives rise to a pair of free spins
which may propagate independently to give rise to spinon excitations.
If the energetic cost of deconfinement is high, the spinons propagate
as a bound pair (confinement) so that the elementary excitation has
spin-1.

\begin{figure}
\begin{center}
\includegraphics[scale=0.7]{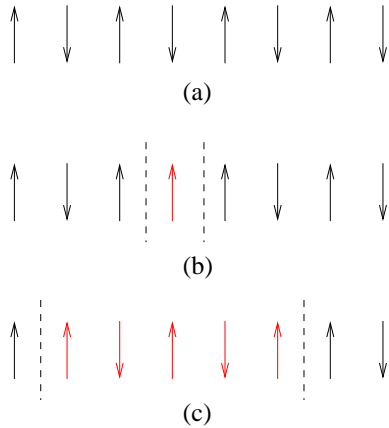}
\end{center}
\caption{(a) N\'{e}el state, (b) a spin flip in the N\'{e}el state creates a pair of
domain walls or spinons the locations of which are shown by dotted
lines, (c) the transverse exchange interaction term in the Hamiltonian
(equation (1)) gives rise to the propagation of independent spinons.}
\end{figure}

Despite considerable effort, there are few experimental evidences
of spinon-like excitations in $d>1$ \cite{key-13,key-14}. One strong candidate
is $Cs_{2}CuCl_{4}$, a spin-$\frac{1}{2}$ Heisenberg antiferromagnet
defined on a spatially anisotropic triangular lattice \cite{key-13}. The
dynamical structure factor $S(k,\omega)$ for $Cs_{2}CuCl_{4}$, where
$k$ and $\omega$ are the momentum and energy transfers in the neutron
scattering experiment, is dominated by a broad continuum which has
been cited as evidence for fractionalized excitations. Kohno et al
\cite{key-15} reanalyzed the neutron scattering data to show that the spinons
are not characteristic of some exotic 2d state but are descendants
of the weakly-coupled excitations of individual chains in the material.
The spectrum also has a sharp dispersing peak attributed to `triplon'
bound states of the spinons. The bound pair lowers its kinetic energy
through propagation between chains. The issue of fractional versus
integer excitations has been extensively investigated in AFM spin-$\frac{1}{2}$
ladders \cite{key-16,key-17,key-18,key-19,key-20}. 
A two-chain ladder consists of two AFM chains
coupled by rung exchange interactions. The spinons of individual chains
are confined even if the rung exchange interaction strength is infinitesimal.
Ladders with strong rung exchange couplings suppress spinon excitations
at all energy scales. Recently, Lake et al \cite{key-21} have carried out
neutron scattering experiments on a weakly-coupled ladder material,
$CaCu_{2}O_{3}$, and shown that deconfined spinons at high energies
evolve into $S=1$ excitations at lower energies. The spinons are
associated with individual chains whereas the $S=1$ excitations are
the triplon excitations, i.e., bound states of spinons. Two approaches
are usually adopted in probing the nature of excitations in spin ladders:
(i) the rung coupling strength dominates and (ii) the rung coupling
strength is weaker than the intra-chain coupling strengths \cite{key-21}.
In this paper, we consider a different case, not studied earlier, in which two $S=\frac{1}{2}$
Ising-like Heisenberg AFM chains, each of which is described by a
Hamiltonian of the type shown in equation (1), are coupled by Ising
or Ising-like Heisenberg AFM exchange interactions. In section II,
we investigate the nature of the low-lying excitations with Ising
rung exchange interactions. In section III, the rung exchange interactions
are considered to be Ising-like Heisenberg AFM in nature. Section
IV contains concluding remarks. 

\section{Ising Rung Exchange Interactions}

We consider an antiferromagnetic two-chain spin ladder with the spins
of magnitude $\frac{1}{2}$. The individual chains of the ladder are
described by the Ising-Heisenberg Hamiltonian (equation(1)). The chains
are coupled by rungs with the corresponding exchange interactions
being of the Ising-type. The ladder Hamiltonian $H_{L}$ is given
by
\small{
\begin{eqnarray}
H_{L} & = &J_{Z}\sum_{\alpha=1}^{2}\sum_{i=1}^{N}S_{i,\alpha}^{Z}S_{i+1,\alpha}^{Z}
+J_{Z}\sum_{i=1}^{N}S_{i,1}^{Z}S_{i,2}^{Z} 
+\frac{J_{XY}}{2}\sum_{\alpha=1}^{2}\sum_{i=1}^{N}
\left(S_{i,\alpha}^{+}S_{i+1,\alpha}^{-}+S_{i,\alpha}^{-}S_{i+1,\alpha}^{+}\right) \\ \nonumber
& = & H_{Z}+H_{XY}
\end{eqnarray}
}
where the index $\alpha=1(2)$ refers to the top (bottom) chain of
the ladder, $i$ denotes the site index and $N$ is the total number
of rungs. We also assume that the anisotropy constant $\epsilon=\frac{J_{XY}}{J_{Z}}$
is $<<1$. Hence, the Ising part of the Hamiltonian, $H_{Z}$, can
be considered to be the unperturbed Hamiltonian with $H_{XY}$, containing
the transverse exchange interactions, providing the perturbation.
Since $H_{Z}$ is AFM in nature, the lowest energy states are the
N\'{e}el states with n.n. spin pairs antiparallel. In the spirit of Villain
\cite{key-6}, we first consider a ladder with an odd number of rungs, i.
e., $N$ = odd and periodic boundary conditions (PBCs). Energies are
measured w. r. t. that of a N\'{e}el configuration of spins. Since there
are $3N$ n.n. spin pairs, the energy of a configuration in which
all such pairs are antiparallel is $E_{N\acute{e}el}=-3N\;\frac{J_{Z}}{4}$.
Since $N$ is odd, a perfect N\'{e}el configuration is not possible and
the lowest energy states of $H_{Z}$ contain a pair of parallel spin
pairs which define the DWs or spinons (figure 2 (a)). The DWs form
a bound pair to ensure minimal energy loss. Any other arrangement
of DWs in the individual chains gives rise to higher energy states.
The lowest energy states are $N$-fold degenerate as there are $N$
possibilities for the location of the bound pair which is an $S_{z}=0$
object. 

\begin{figure}
\begin{center}
\includegraphics[scale=0.7]{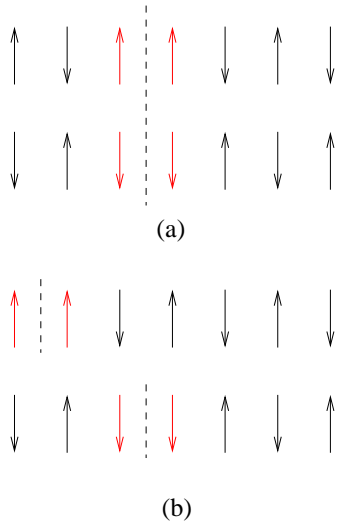}
\end{center}
\caption{(a) A pair of DWs in the minimum energy configuration of a ladder
with an odd number $N$ of rungs. (b) Motion of the DW in the top
chain leaves in its wake ferromagnetically aligned rung spins which
raises the energy of the system. }
\end{figure}

We next consider the effect of the perturbing Hamiltonian on the minimum
energy states. The transverse exchange interaction can give rise to
independent DW motion in the chains which, however, costs energy as
a propagating DW leaves in its wake ferromagnetically aligned rung
spins (figure 2(b)). The energy cost increases as the distance between
the DWs increases resulting in confinement of the DW pair. We investigate
the dynamics of the DW pair perturbatively using a low-energy effective
Hamiltonian (LEH) \cite{key-22,key-23}. The $N$-fold degenerate DW pair states
$|p_{i}\rangle,\; i=1,...,N$ (figure 2(a)) constitute the low-energy
manifold and have energy $E_{0}=-\frac{3NJ_{Z}}{4}+J_{Z}$. The higher
energy states of $H_{Z}$ are denoted by $|q_{\alpha}\rangle$ with
energy $E_{\alpha}$. The perturbing Hamiltonian $H_{XY}$ connects
the low-energy manifold to the manifold of higher-energy states. In
general, perturbation lifts the degeneracy of the low-energy manifold
leading to an effective Hamiltonian operating in the space of states
associated with the low-energy manifold. Diagonalization of the $n$-th
order $(n=1,2,...)$ effective Hamiltonian in the low-energy subspace
of states reproduces the $n$-th order energy corrections to the low-energy
unperturbed states. Using degenerate perturbation theory, the first
order LEH is given, up to an overall constant, by \cite{key-23}
\small{
\begin{eqnarray}
H_{eff}^{(1)}=\sum_{ij}|p_{i}\rangle\langle p_{i}|H_{XY}|p_{j}\rangle\langle p_{j}|
\end{eqnarray}
}
The second-order LEH has the form 
\small{
\begin{eqnarray}
H_{eff}^{(2)}=\sum_{ij}\sum_{\alpha}
|p_{i}\rangle\frac{\langle p_{i}|H_{XY}|q_{\alpha}\rangle\langle q_{\alpha}|H_{XY}|p_{j}\rangle}
{E_{0}-E_{\alpha}}\langle p_{j}|
\end{eqnarray}
}
Since the matrix element $\langle p_{i}|H_{XY}|p_{j}\rangle=0$, the
LEH is determined using the second-order expression in equation (4).
Two types of processes contribute to $H_{eff}^{(2)}$. In {}``diagonal''
processes, the spins in an antiparallel pair exchange and then reexchange
back to the original configuration $(|p_{i}\rangle=|p_{j}\rangle)$.
Such processes do not lift the degeneracy and give rise to a constant
energy shift. We neglect this contribution in deriving the effective
Hamiltonian. In the off-diagonal processes, the spins in two antiparallel
pairs belonging to the four-spin plaquettes bordering the bound DW
pair are interchanged. For the DW pair state shown in figure 2(a),
the intermediate states $|q_{i}\rangle,\; i=1,..,4$ are given by 
\begin{eqnarray}
|q_{1}\rangle & = & \begin{array}{ccccccc}
\uparrow & \Uparrow & \Downarrow & \uparrow & \downarrow & \uparrow & \downarrow\\
\downarrow & \uparrow & \downarrow & \downarrow & \uparrow & \downarrow & \uparrow\end{array}\\ \nonumber
|q_{2}\rangle & = & \begin{array}{ccccccc}
\uparrow & \downarrow & \uparrow & \uparrow & \downarrow & \uparrow & \downarrow\\
\downarrow & \Downarrow & \Uparrow & \downarrow & \uparrow & \downarrow & \uparrow\end{array}\\ \nonumber
|q_{3}\rangle & = & \begin{array}{ccccccc}
\uparrow & \downarrow & \uparrow & \Downarrow & \Uparrow & \uparrow & \downarrow\\
\downarrow & \uparrow & \downarrow & \downarrow & \uparrow & \downarrow & \downarrow\end{array}\\ \nonumber
|q_{4}\rangle & = & \begin{array}{ccccccc}
\uparrow & \downarrow & \uparrow & \uparrow & \downarrow & \uparrow & \downarrow\\
\downarrow & \uparrow & \downarrow & \Uparrow & \Downarrow & \downarrow & \uparrow\end{array}
\end{eqnarray}
\begin{figure}
\begin{center}
\includegraphics[scale=0.5]{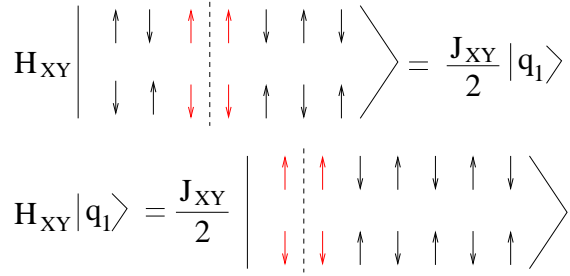}
\end{center}
\caption{Two successive applications of $H_{XY}$ on low-energy states shifts
the location of the bound DW pair by two lattice constants.}
\end{figure}
The spin pairs deviated from the arrangement shown in figure 2 (a)
are marked by double arrows. The perturbing Hamiltonian, $H_{XY}$,
acting on the intermediate $|q_{\alpha}\rangle$ states shifts the
location of the bound DW pair by two lattice constants either towards
the left or the right. This is illustrated in figure 3 for the state
$|q_{1}\rangle$. The energy of the states $|q_{\alpha}\rangle\;(\alpha=1,...,4)$
is $E_{\alpha}=-\frac{3NJ_{Z}}{4}+2J_{Z}$. The second-order LEH is
thus given by 
\small{
\begin{eqnarray}
H_{eff}^{(2)}=-\frac{\epsilon^{2}J_{z}}{2}
\sum_{i}\left(|p_{i+2}\rangle\langle p_{i}|+|p_{i-2}\rangle\langle p_{i}|\right)
\end{eqnarray}
}
where $\epsilon=\frac{J_{XY}}{J_{Z}}$. The off-diagonal processes
are equivalent to {}``ring'' exchanges involving four spins. The
full second-order Hamiltonian, defined in the low-energy manifold,
is thus given by
\small{
\begin{eqnarray}
H_{eff}=H_{Z}+H_{ring}
\end{eqnarray}
}
where 
\small{
\begin{eqnarray}
H_{ring}=J_{ring}\sum_{\square}\left(S_{1}^{+}S_{2}^{-}S_{3}^{+}S_{4}^{-}+h.c.\right)
\end{eqnarray}
}
with $J_{ring}=-\frac{\epsilon^{2}J_{Z}}{2}$ and the sum over all
elementary plaquettes of the ladder. Ring or cyclic exchange interactions
(equation (8)) also appear in the perturbative effective Hamiltonian
theories developed for the XXZ Heisenberg model on the checkerboard
lattice \cite{key-24} and in the case of an easy-axis Kagom\'{e} antiferromagnet
\cite{key-25}. In the ladder model, the ring exchange interaction has the
effect of deconfining the bound DW pair. In the low-energy subspace,
the dispersion relation of the bound pair can be determined in a straightforward
manner. The eigenstate $\psi(k)$ of the pair can be written as a
linear combination of staes $|j\rangle(j=1,2,...,N)$ where $j$ denotes
the location of the bound DW pair. 
\small{
\begin{eqnarray}
|\psi(k)\rangle=\frac{1}{\sqrt{N}}\sum_{j=1}^{N}e^{ikj}|j\rangle
\end{eqnarray}
}
$H_{eff}$ (equation (7)) operating on $|\psi(k)\rangle$ yields the
eigenvalue
\small{
\begin{eqnarray}
\omega_{b}(k)=J_{Z}(1-\epsilon^{2}\cos2k)
\end{eqnarray}
}
The subscript `$b$' in $\omega_{b}$ denotes that the dispersion
relation is that of a bound DW pair. 

We next consider a two-chain spin ladder with an even number $N$
of rungs and described by the Hamiltonian in equation (7) satisfying
PBCs. The low-lying excitation spectrum is obtained in the subspace
of degenerate eigenstates of $H_{Z}$ which are generated by flipping
all the spins in a block of $\mu$ ($\mu$ may be odd/even) adjacent
rungs in the ground state (N\'{e}el state) of $H_{Z}$. Each of the states
contains two bound DW pairs (figure 4) and has energy 
\small{
\begin{eqnarray}
E_{DW}=-\frac{3NJ_{Z}}{4}+2J_{Z}
\end{eqnarray}
}
\begin{figure}
\begin{center}
\includegraphics[scale=0.7]{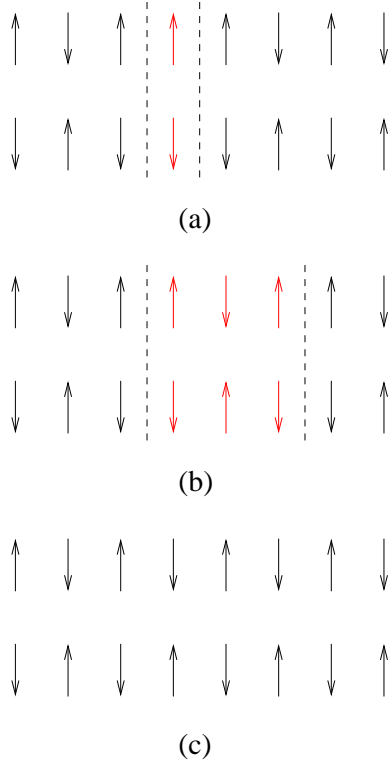}
\end{center}
\caption{Bound DW pair states for (a) $\mu=1$ and (b) $\mu=3$. The dotted
lines indicate the locations of the bound DW pairs; (c) One of the
ground states, $\psi_{N\acute{e}el\;1}$, of $H_{Z}$.}
\end{figure}
The $z$-component of the total spin of each state, $S_{z}^{tot}=0$.
We consider $\mu$ to be odd with the degenerate eigenstates of $H_{Z}$
given by 
\small{
\begin{eqnarray}
\psi_{1}(k) & = & \sqrt{\frac{2}{N}}\sum_{j}e^{ikj}
\mathcal{S}_{j}\psi_{N\acute{e}el1}\\ \nonumber
\psi_{3}(k) & = & \sqrt{\frac{2}{N}}\sum_{j}e^{ikj}
\mathcal{S}_{j}\mathcal{S}^{\prime}_{j+1}\mathcal{S}_{j+2}\psi_{N\acute{e}el1} \\ \nonumber
& ... & \\ \nonumber
\psi_{N-1}(k) & = & \sqrt{\frac{2}{N}}\sum_{j}e^{ikj}
\mathcal{S}_{j}\Pi_{\mu=1}^{\frac{N}{2}-1}\mathcal{S}^{\prime}_{j+2\mu-1}
\mathcal{S}_{j+2\mu}\psi_{N\acute{e}el1}
\end{eqnarray}
}
\noindent where $\mathcal{S}_{k}=S_{k,1}^{+}S_{k,2}^{-}$ and 
$\mathcal{S}^{\prime}_{k}=S_{k,1}^{-}S_{k,2}^{+}$.
$\psi_{N\acute{e}el1}$ is the ground state of $H_{Z}$ shown in
figure 4(c). The Hamiltonian $H_{ring}$ in (7) has the following
matrix elements between the Ising eigenstates :
\small{
\begin{eqnarray}
\langle\psi_{1}(k)|H_{ring}|\psi_{3}(k)\rangle & = & \langle\psi_{3}(k)|H_{ring}|\psi_{5}(k)\rangle
= ...=J_{ring}\left(1+e^{-2ik}\right) \equiv v
\end{eqnarray}
}
where $J_{ring}=-\frac{\epsilon^{2}J_{Z}}{2}$. The low-lying excited
state of the Hamiltonian $H_{eff}$ (equation (7)) is given by 
\small{
\begin{eqnarray}
\psi_{DW}(k)=\sum_{\nu=1}^{N/2}C_{\nu}\psi_{2\nu-1}(k)
\end{eqnarray}
}
From the eigenvalue equation $H_{eff}\psi_{DW}(k)=\lambda\psi_{DW}(k)$,
one obtains
\small{
\begin{eqnarray}
\sum_{\nu^{\prime}=1}^{N/2}\langle\nu|H_{eff}|\nu^{\prime}\rangle C_{\nu^{\prime}}=\lambda C_{\nu}
\end{eqnarray}
}
where
\small{
\begin{eqnarray}
\langle\nu|H_{eff}|\nu^{\prime}\rangle & = & 2J_{Z}\;\; for\;\nu^{\prime}=\nu \\ \nonumber
& = & v\;\; for\;\nu^{\prime}=\nu+1 \\ \nonumber
& = & v^{*}\;\; for\;\nu^{\prime}=\nu-1 \\ \nonumber
& = & 0\;\; otherwise
\end{eqnarray}
}
The diagonal matrix element is $2J_{Z}$ as energies are measured
with respect to the N\'{e}el state energy $-\frac{3NJ_{Z}}{4}$ (see equation
(11)). We choose the coefficients $C_{\nu}$'s to be $C_{\nu}=e^{-i\phi\nu}$.
The eigenvalues constitute an excitation continuum given by 
\small{
\begin{eqnarray}
\lambda=2J_{Z}\left[1-\epsilon^{2}\cos k\cos(k+\phi)\right]
\end{eqnarray}
}
where $-\pi<\phi\leq\pi$.

\begin{figure}
\begin{center}
\includegraphics[scale=0.7]{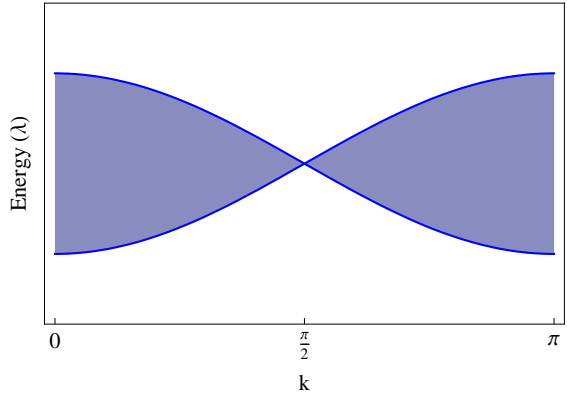}
\end{center}
\caption{Excitation continuum with energies given in equation (17) and $\epsilon=0.15$}
\end{figure}

Figure 5 shows the excitation continuum with upper and lower bounds
given by $2J_{Z}(1\pm\epsilon^{2}\cos k)$. The degenerate eigenstates
of $H_{Z}$ defined in (12) correspond to the top chain of the spin
ladder being in as $S_{tot,1}^{z}=+1$ and the bottom chain being
in an $S_{tot,2}^{z}=-1$ state. One can construct a set of degenerate
eigenstates with the situation reversed. Also, $\mu$, the number
of adjacent rungs constituting the block of flipped spins can be even
$(\mu=2,4,6,...etc.)$. There are two distinct sets of such states
\cite{key-8,key-26}. All these subspaces of states give rise to the same excitation
continuum (figure 5). The excitation continuum arises due to the motion
of two walls each of which consists of a bound pair of DWs.

One notes that the effects of $H_{ring}$, for the two-chain ladder,
and $H_{XY}$, for the 1d chain, on the DW states are similar. In
the first case, a bound pair of DWs shifts by two lattice constants
and in the second case a single DW shifts by the same distance. In
the case of a single chain, the eigenvalues $\lambda_{1d}$ constituting
the excitation continuum are 
\small{
\begin{eqnarray}
\lambda_{1d}=J_{Z}\left[1+2\epsilon\cos k\cos(k+\phi)\right]
\end{eqnarray}
}
Comparing equations (17) and (18), one finds that the spread of the
continuum around the unperturbed level is less in the case of the
spin ladder. 

The dynamic form factor, $S_{11}(q,\omega)$, associated with the
bound DW pair is defined at $T=0$ by 
\small{
\begin{eqnarray}
S_{11}(q,\omega)=\sum_{f}|\langle f|A(q)|g\rangle|^{2}\delta(\omega-E_{f}+E_{g})
\end{eqnarray}
}
where
\small{
\begin{eqnarray}
A(q)  =  \frac{1}{\sqrt{N}}\sum_{l}A_{l}e^{iql} 
A_{l}  =  \left(S_{l,1}^{+}S_{l,2}^{-}+S_{l,1}^{-}S_{l,2}^{+}\right)
\end{eqnarray}
}
In (19), $|g\rangle$ and $|f\rangle$ are the ground and excited
states of $H_{eff}$ connected by $A(q)$, with energies $E_{g}$
and $E_{f}$ respectively, $\omega$ and $q$ are the frequency and
wave number of the excitation. $S_{11}(q,\omega)$, involving a pair
of spin deviations, could be probed by the light-scattering techniques
\cite{key-27,key-28}. Upto the first order of $J_{ring}$ ($J_{ring}<<J_{Z}$
in (7)), 
\small{
\begin{eqnarray}
|g\rangle\simeq\psi_{N\acute{e}el\;\;1}+\frac{1}{E_{0}-H_{Z}}H_{ring}\psi_{N\acute{e}el1}
\end{eqnarray}
}
where $E_{0}$ is the energy of $\psi_{N\acute{e}el1}$. Since $H_{ring}$
acting on $\psi_{N\acute{e}el1}$ creates two bound DW pairs, $\frac{1}{E_{0}-H_{Z}}=-\frac{1}{2J_{Z}}.$Thus,
\small{
\begin{eqnarray}
A(q)|g\rangle  \simeq \frac{1}{\sqrt{2}}
\left(1+\frac{\epsilon^{2}}{2}\cos q\right)\psi_{1}(q)
 +\frac{1}{\sqrt{2}}\frac{v^{*}}{-2J_{Z}}\psi_{3}(q)
\end{eqnarray}
}
where $\psi_{1}(q)$ and $\psi_{3}(q)$ are as defined in equation
(12) and $v^{*}=-\frac{\epsilon^{2}J_{Z}}{2}\left(1+e^{2iq}\right)$.
Using equation (22) and (19) and the expression (14) for $|f\rangle=\psi_{DW}(k)$,
one gets following the procedures described in \cite{key-7,key-26}
\small{
\begin{eqnarray}
S_{11}(q,\omega)
& \simeq & \frac{\sqrt{4|v|^{2}-\Omega^{2}}}{2\pi|v|^{2}}\left(1+\epsilon^{2}\cos q-\frac{\Omega}{J_{Z}}\right)for|\Omega|<2|v| \nonumber \\  
& = & 0\; otherwise 
\end{eqnarray}
}
with $\Omega=\omega-2J_{Z}$. 

The expression (23) is similar to that for the dynamic structure factor
$S_{xx}(q,\omega)$ of the $1d$ chain obtained in first order perturbation
theory \cite{key-7} ($A_{l}=S_{l}^{x}$ in equation (20)) except that in
the latter case, $\Omega=\omega-J_{Z}$ and the contribution of the
anisotropy term is to first order in $\epsilon$. Figure 6 shows the
plots of $S_{11}(q,\omega)\times2J_{Z}|\cos q|$ versus $\frac{\omega}{J_{Z}}$
for $\epsilon=0.15$ and for various values of the wave number $q$.
The lineshape is almost symmetric in contrast to the prominent asymmetry
found in the 1d case \cite{key-7}.

\begin{figure}
\begin{center}
\includegraphics[scale=0.6]{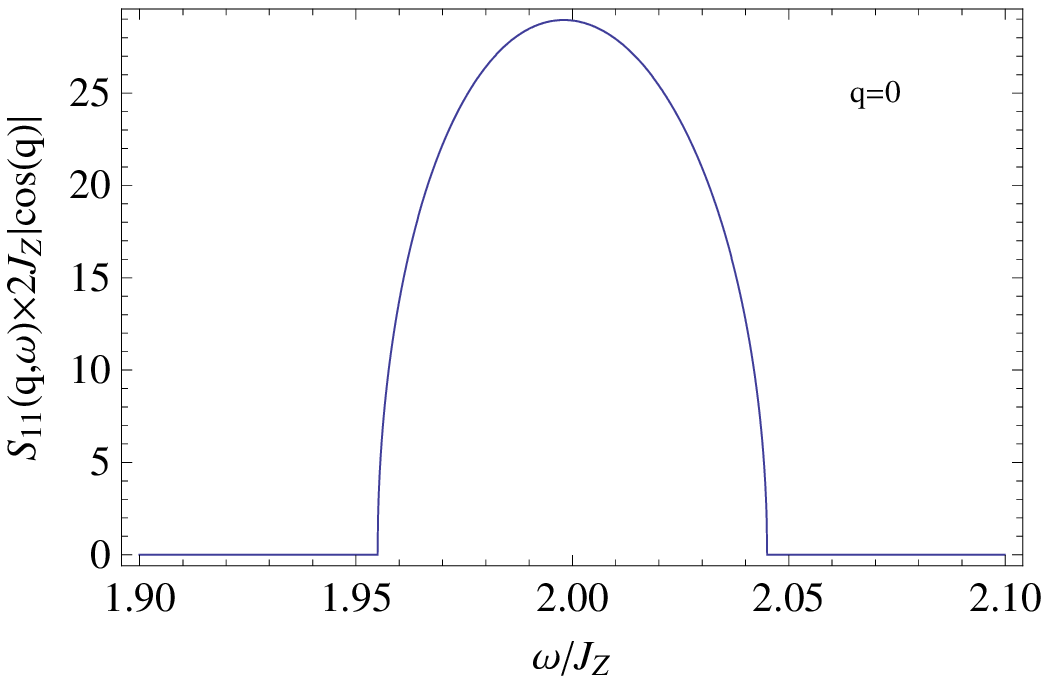}
\includegraphics[scale=0.6]{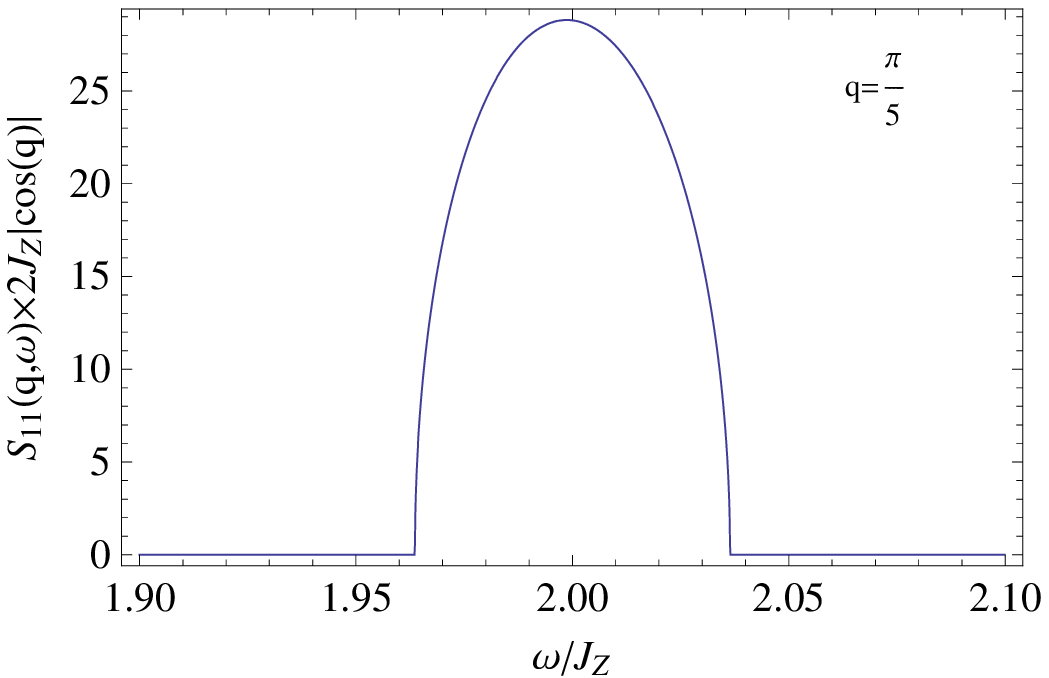}\linebreak
\includegraphics[scale=0.6]{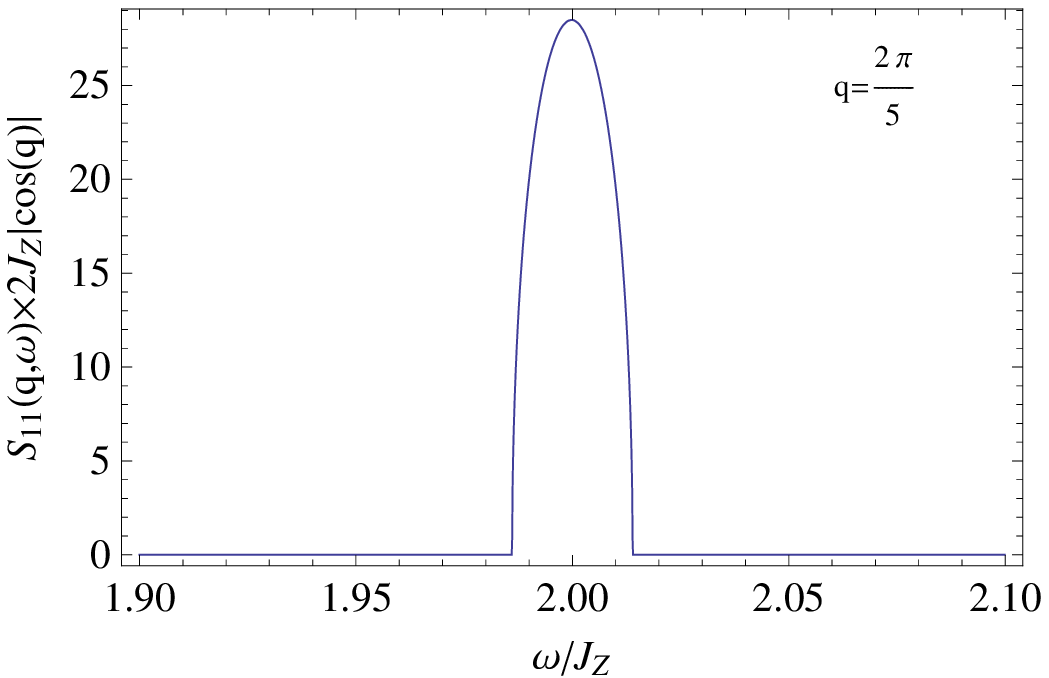}
\includegraphics[scale=0.6]{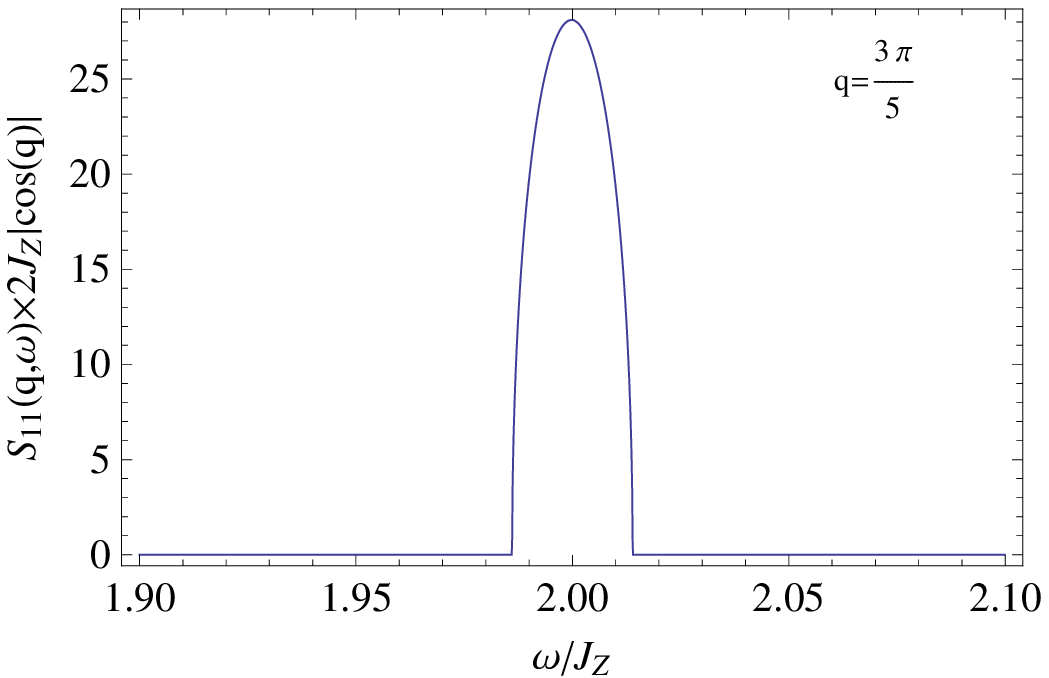}\linebreak
\includegraphics[scale=0.6]{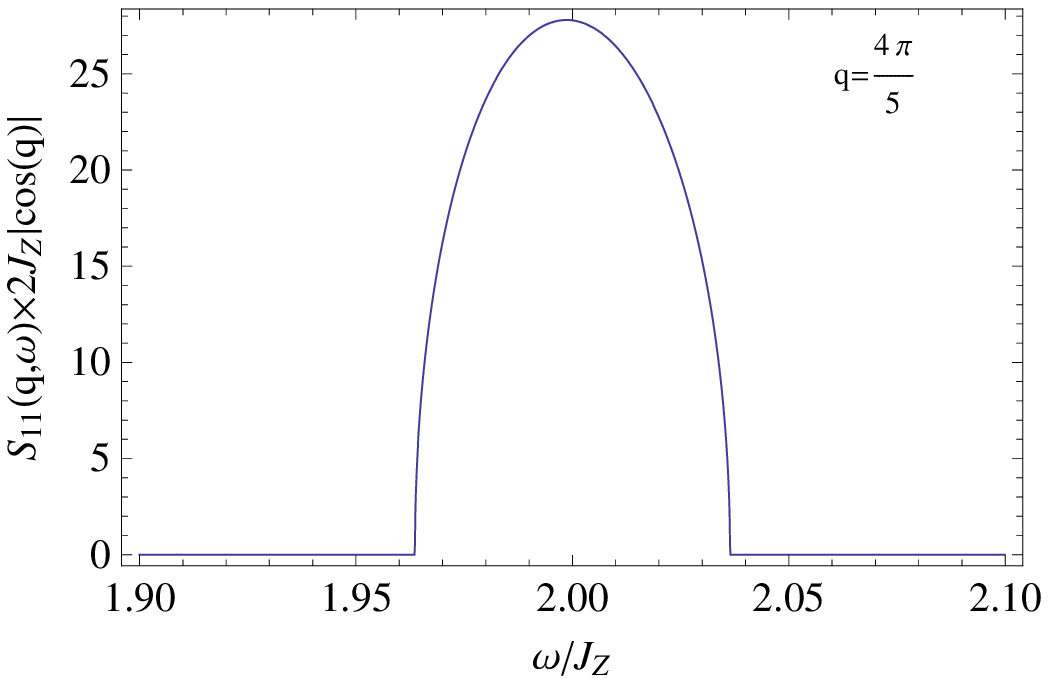}
\includegraphics[scale=0.6]{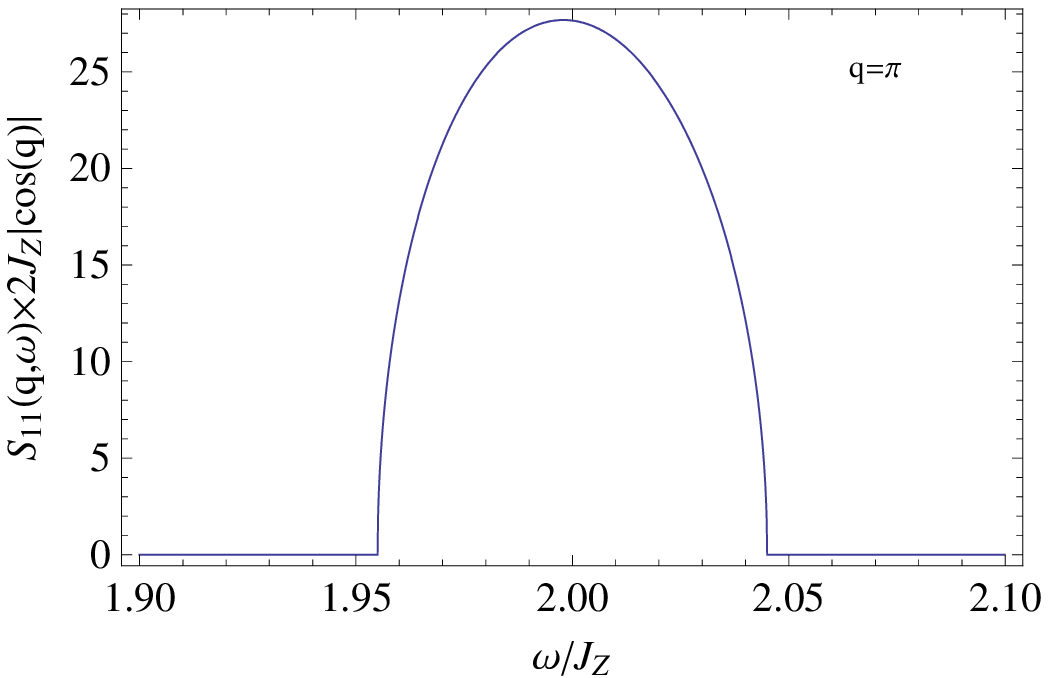}
\end{center}
\caption{$S_{11}(q,\omega)\times2J_{Z}|\cos q|$ evaluated in second-order
perturbation theory for $\epsilon=0.15$ versus $\frac{\omega}{J_{Z}}$
for different values of $q$. }
\end{figure}
\section{Ising-Heisenberg Rung Exchange Interactions}

We now consider the case in which the rung exchange interactions of
the two-chain spin ladder are Ising-like Heisenberg-type. The Hamiltonian
is given by 
\small{
\begin{eqnarray}
H_{R} & = & J_{Z}\sum_{\alpha=1}^{2}\sum_{i=1}^{N}S_{i,\alpha}^{z}S_{i+1,\alpha}^{z}
+J_{Z}\sum_{i=1}^{N}S_{i,1}^{z}S_{i,2}^{z}
+\frac{J_{XY}}{2}\sum_{\alpha=1}^{2}\sum_{i=1}^{N}
\left(S_{i,\alpha}^{+}S_{i+1,\alpha}^{-}+S_{i,\alpha}^{-}S_{i+1,\alpha}^{+}\right) \\ \nonumber
& & +\frac{J_{XY}}{2}\sum_{i=1}^{N}\left(S_{i,1}^{+}S_{i,2}^{-}+S_{i,1}^{-}S_{i,2}^{+}\right) \\ \nonumber
& = & H_{Z}+H_{XY}
\end{eqnarray}
}
The Hamiltonian (24) differs from $H_{L}$ is equation (2) by the
addition of the last term. The ground states of $H_{Z}$ are the doubly
degenerate N\'{e}el states. We consider the N\'{e}el state $\psi_{N\acute{e}el1}$
shown in figure 4(c). The ladder can be divided into two sublattices
$A$ and $B$ such that in $\psi_{N\acute{e}el1}$ the $A(B)$ sublattice
spins are pointing up (down). The ground state energy $E_{0}=-\frac{3NJ_{Z}}{4}$.

The lowest energy excitation of the unperturbed Hamiltonian is obtained
by flipping a single spin in either the $A\;(S_{tot}^{z}=-1)$ or
the $B\;(S_{tot}^{z}=+1)$ sublattice. We consider the latter case
with $|i\rangle$ denoting the state in which the flipped spin is
located in the $i$th rung (figure 7(a)). These excited states are
$N$-fold degenerate with the energy 
\small{
\begin{eqnarray}
E_{1}=-\frac{3NJ_{Z}}{4}+\frac{3J_{Z}}{2}
\end{eqnarray}
}
The perturbing Hamiltonian acting on the state $|i\rangle$ generates
the following states 
\small{
\begin{eqnarray}
H_{XY}|i\rangle=\frac{\epsilon J_{Z}}{2}\left[|1\rangle+|2\rangle+|3\rangle+|4\rangle+....\right]
\end{eqnarray}
}
where $\epsilon=\frac{J_{XY}}{J_{Z}}$. The states $|m\rangle(m=1,...,4)$
are shown in figure 7(b) with energy 
\small{
\begin{eqnarray}
E_{m}=-\frac{3NJ_{Z}}{4}+\frac{5J_{Z}}{2}
\end{eqnarray}
}
The other states which are generated when $H_{XY}$ acts on the state
$|i\rangle$ have higher energies and are hence not considered. $H_{XY}$
acting on the states $|m\rangle$ gives 
\small{
\begin{eqnarray}
H_{XY}|m\rangle & = & \frac{\epsilon J_{Z}}{2}\left(|i\rangle+|i-1\rangle\right),\; m=1,2\\ \nonumber
 & = & \frac{\epsilon J_{Z}}{2}\left(|i\rangle+|i+1\rangle\right),\; m=3,4
\end{eqnarray}
}

In first order perturbation theory there is no energy correction.
A finite energy correction is obtained in the second order perturbation
theory. The effective LEH (equation (4)) with $|p_{i}\rangle=|i\rangle,\;|p_{j}\rangle=|j\rangle,\;|q_{\alpha}\rangle=|m\rangle$
and $E_{\alpha}=E_{m}$ in the same order is given by 
\small{
\begin{eqnarray}
H_{eff1}^{(2)} & = & -\frac{\epsilon^{2}J_{Z}}{2}\sum_{i}
\left(|i\rangle\langle i-1|+|i\rangle\langle i+1|\right)\\
 & = & -\frac{\epsilon^{2}J_{Z}}{2}\sum_{i=1}^{N}\sum_{\delta=-1,1}
\left(S_{i,1}^{+}S_{i+\delta,2}^{-}+S_{i,1}^{-}S_{i+\delta,2}^{+}\right)
\end{eqnarray}
}
The effect of this Hamiltonian on the low-energy excited state $|i\rangle$
(figure 7(a)) is to shift the flipped spin from the $i$th to the
$(i+1)$th or the $(i-1)$th rungs. Since the flipped spins are located
in the $B$ sublattice, the shift is in the diagonal direction. The
flipped spin is associated with a bound pair of DWs in a chain. The
bound pair lowers its kinetic energy by propagating between chains.
The full second-order Hamiltonian defined in the low-energy manifold
of states with single spin flips, is 
\small{
\begin{eqnarray}
H_{eff1}=H_{Z}+H_{eff1}^{(2)}
\end{eqnarray}
}
As before, we have not included the terms arising from the {}``diagonal''
processes in equation (4) as they give rise to a constant energy shift.
The low energy excited state with $S_{tot}^{z}=+1$ can be constructed
as 
\small{
\begin{eqnarray}
|k\rangle=\frac{1}{\sqrt{N}}\sum_{j=1}^{N}e^{ikj}S_{j,B}^{+}|\psi_{N\acute{e}el1}\rangle
\end{eqnarray}
}
where `$B$' denotes the $B$ sublattice. The dispersion relation
for the propagation of the flipped spin or equivalently the bound
DW pair is given by 
\small{
\begin{eqnarray}
E_{1}(k)=J_{Z}\left(1-\epsilon^{2}\cos k\right)
\end{eqnarray}
}
where the energy is measured w. r. t the N\'{e}el state energy. The bound
DW pair moves diagonally across the spin ladder. The unperturbed Hamiltonian,
$H_{Z}$, is the same irrespective of whether the rung exchange interactions
are Ising-type or Ising-Heisenberg-type. Thus, the lowest unperturbed
excited state is the single flip state in both the cases. The excitation
has a localized character when the rung exchange interactions are
of the Ising-type. The bound DW pair associated with the single spin-flip
can not propagate between the chains as the inter-chain interactions
are Ising-like and propagation of the DWs in a single chain is, as
pointed out before, energetically prohibitive. The energy, $-\frac{3J_{Z}}{2}$,
of the localized excitation is lower than that of the propagating
excitations involving two bound DW pairs when the rung exchange interactions
are of the Ising-type (section 2). In this case, propagating excitations
with the lowest energy involve two bound DW pairs rather than one. 

\begin{figure}
\begin{center}
\includegraphics[scale=0.6]{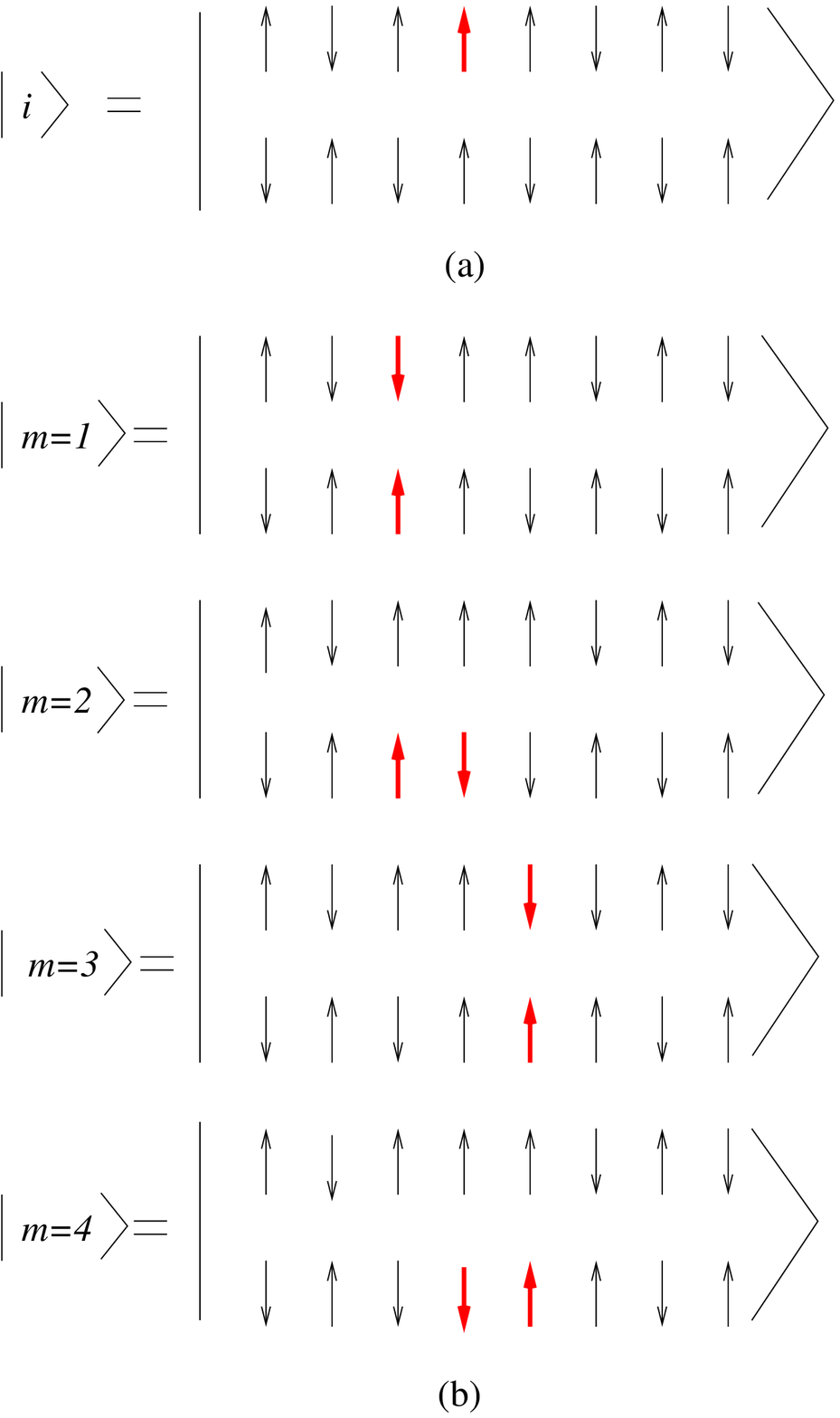}
\end{center}
\caption{(a) The lowest energy excitation $(S_{tot}^{z}=+1)$ of the unperturbed
Hamiltonian $H_{Z}$ in equation (24). The deviated spin is represented
by a thick arrow. (b) The perturbing Hamiltonian, $H_{XY}$, acting
on the state $|i\rangle$ generates the states $|m\rangle(m=1,2,3,4).$
The spin deviations from the state $|i\rangle$ are shown by thick
arrows. }
\end{figure}

\section{Concluding Remarks}
AFM spin models in which the existence of spinons is well-established
include the spin-$\frac{1}{2}$ Heisenberg AFM chain \cite{key-1}, the
Majumdar-Ghosh model \cite{key-29} and the Haldane-Shastry model \cite{key-30,key-31}.
The physical picture of a spinon as a DW between two degenerate ground
states emerges in the Ising-Heisenberg limit of the AFM Hamiltonian
\cite{key-1,key-7,key-8}. In the case of the MG model, the spin-$\frac{1}{2}$
excitation acts as a DW between the two dimerized ground states of
the model \cite{key-32,key-33}. In a closed chain, the DWs occur in pairs
so that the lowest-lying excitation is given by the two-spinon continuum.
The spinons are deconfined in this case and can move away from each
other. There is no energy cost in moving the spinons far apart. This
is not so when two AFM chains are coupled in the form of a spin ladder.
Let $J_{\bot}$and $J_{||}$ be the strengths of the rung and intra-chain
n.n. exchange interactions respectively. The spinon excitations of
individual chains are confined by even an infinitesimal coupling strength
$J_{\bot}$\cite{key-18,key-21}. The two $S=\frac{1}{2}$ spinons form a bound
state giving rise to singlet and triplet excitation branches. In the
case of the strongly coupled ladder $(J_{\bot}>>J_{||})$, the elementary
excitation is a triplet. Lake et al. \cite{key-21} carried out neutron
scattering experiments on the weakly-coupled $(J_{\bot}<<J_{||})$
ladder material $CaCu_{2}O_{3}$ and obtained evidence of the singlet
excitation mode. The spinon continuum was observed at high energies
for which the chains are effectively decoupled. The spinons in a chain
evolve into an $S=1$ excitation at lower energies thereby confirming
that the $S=1$ {}``triplon'' excitation is a bound state of two
spinons and not a conventional magnon. 

In this paper, we study a two-chain $S=\frac{1}{2}$ AFM spin ladder
in which the individual chains are described by the Ising-like Heisenberg
Hamiltonian and the rung couplings are of the Ising-type. Using a
low-energy effective Hamiltonian approach, we establish that in a
ladder with an odd number of rungs the spinons (DWs) form a bound
pair. A four-spin ring exchange interaction in the effective Hamiltonian
is responsible for the delocalization of the bound pair. In the case
of a ladder with an even number of rungs, the low-lying propagating
excitation involves two bound pairs of DWs which can move away from
each other giving rise to a continuum of excitations. The physical
origin of the excitation continuum is similar to that in the case
of the Ising-Heisenberg AFM chain in 1d except that in the former
case the spinons form bound pairs and the dispersion of the excitation spectrum 
is a higher-order effect
in perturbation theory. This results in an almost symmetric lineshape
in the case of the dynamic structure factor $S_{11}(q,\omega)$ (Figure
6) in contrast to the asymmetry observed in the structure factor $S_{xx}(q,\omega)$ in
1d \cite{key-7}. The delocalization of a bound spinon pair
is brought about via a ring or a diagonal exchange interaction term
in the effective Hamiltonian.

We further consider a second model in which the rung exchange interactions
are described by the Ising-like Heisenberg Hamiltonian. In this case
also, the spinon pair in a single chain is bound and the bound pair
lowers its kinetic energy by hopping between chains. Kohno et al.
\cite{key-15} have studied a $S=\frac{1}{2}$ spatially anisotropic frustrated
Heisenberg antiferromagnet in 2d in the weak interchain coupling regime.
The model provides a good quantitative fit to the inelastic inelastic
neutron scattering data of the triangular antiferromagnet $Cs_{2}CuCl_{4}.$
The spectrum consists of a continuum arising from the deconfinement
of spinons in individual chains and a sharp dispersing peak associated
with the coherent propagation of a triplon bound state of two spinons
between neighbouring chains. In the case of our model, the bound pair
has $S^{z}=+1$. One can similarly construct an $S^{z}=-1$ excitation.
In summary, we have studied ladder models with the Ising-like Heisenberg
Hamiltonian describing the interactions in individual chains. The
rung interactions may be pure Ising or Ising-like Heisenberg.
The ladder models studied in this paper
share some common features with models in which the exchange interactions
are isotropic. 
In the latter case, two types of models have generally been considered : ladder models 
in which the rung exchange interactions are the most dominant and models which describe 
spin chains coupled by weak exchange interactions. The models considered in this paper 
belong to a category not studied earlier and provide considerable physical insight on the 
origin of spinon confinement and how the bound spinon pairs delocalize. 
The isotropic models have, however, a richer dynamics with interactions
generating cascades of virtual particles so that the two-body confinement
problem becomes a many-body one \cite{key-21}. The major common feature
emerging from the study of ladder models with both Ising-like and
isotropic exchange interactions appears to be the confinement of spinons
in the form of bound states. The origin of excitation continua in
specific cases lies in multi-triplon excitations rather than in the fractionalization
of excitations \cite{key-18}.

\end{document}